\documentclass{IEEEcsmag}

\usepackage[colorlinks,urlcolor=blue,linkcolor=blue,citecolor=blue]{hyperref}
\usepackage{booktabs}
\usepackage{upmath}
\usepackage{graphics}
\usepackage{comment} 
\usepackage{subcaption}
\usepackage{marginnote}
\usepackage{tablefootnote}
\usepackage{multirow}

\jvol{XX}
\jnum{XX}
\paper{8}
\jmonth{May/June}
\jname{IT Professional}
\pubyear{2019}

\begin{document}

\sptitle{Department: Head}
\editor{Editor: Name, xxxx@email}

\title{Reproducibility of the First Image of a Black Hole in the Galaxy M87 from the Event  Horizon  Telescope (EHT) Collaboration
}

\author{\IEEEauthorblockN{
R. Patel\IEEEauthorrefmark{1}, 
B. Roachell\IEEEauthorrefmark{1}, 
S. Ca\'ino-Lores\IEEEauthorrefmark{1}, 
R. Ketron\IEEEauthorrefmark{1}, 
J. Leonard\IEEEauthorrefmark{1}, 
N. Tan\IEEEauthorrefmark{1}, 
K. Vahi\IEEEauthorrefmark{2}, 
D.~A.~Brown\IEEEauthorrefmark{3}, 
E. Deelman\IEEEauthorrefmark{2}, and 
M. Taufer\IEEEauthorrefmark{1}}
\IEEEauthorblockA{U. Tennessee - Knoxville\IEEEauthorrefmark{1}, U. Southern California\IEEEauthorrefmark{2}, and Syracuse U.\IEEEauthorrefmark{3}\\
Emails: \IEEEauthorrefmark{1}taufer@utk.edu,
\IEEEauthorrefmark{2}deelman@isi.edu,
\IEEEauthorrefmark{3}dabrown@syr.edu}}

\markboth{Department Head}{Paper title}

\begin{abstract}

This paper presents an interdisciplinary effort aiming to develop and share sustainable knowledge necessary to analyze, understand, and use published scientific results to advance reproducibility in multi-messenger astrophysics. Specifically, we target the breakthrough work associated with the generation of the first image of a black hole, called M87. The image was computed by the Event Horizon Telescope Collaboration. Based on the artifacts made available by EHT, we deliver documentation, code, and a computational environment to reproduce the first image of a black hole. Our deliverables support new discovery in multi-messenger astrophysics by providing all the necessary tools for generalizing methods and findings from the EHT use case. Challenges encountered during the reproducibility of EHT results are reported. The result of our effort is an open-source, containerized software package that enables the public to reproduce the first image of a black hole in the galaxy M87.
\end{abstract}

\maketitle


\section{Introduction}

Developing reproducible analyses is a challenging aspect of scientific research. Few real-world studies have been performed to provide guidance on the necessary processes and products, especially in domains relying on scientific computing. There reproducibility is limited by the availability of data, software, platforms, and documentation. Consequently, despite a group's best efforts, other scientists attempting to reproduce an analysis may find that the necessary information is incomplete. 

We present an interdisciplinary effort to develop and share sustainable knowledge necessary to understand, reproduce, and reuse the published scientific results of the Event Horizon Telescope (EHT) project's analysis of the black hole in the center of the M87 galaxy~\cite{akiyama2019firstVI}. Unlike our previous reproduction of Advanced LIGO's observations~\cite{brown2021reproducing}, none of the authors of this paper was involved in the original EHT analysis. Thus, our work builds exclusively on the several papers describing the EHT project workflow~\cite{akiyama2019firstIV}, data~\cite{akiyama2019firstIII,eht-dataset}, and software~\cite{eht-code} that are available online. Each EHT paper presents specific aspects of the scientific discovery but a comprehensive approach including documentation, software, and environment to reproduce the published results of the EHT project is still missing. To this end, this paper follows rigorous reproducibility directions and expands preliminary work presented in a poster~\cite{ketron2021case}.

As part of our contributions, we investigate the availability and integrity of the data used to recreate the images of the M87 black hole. We model the image processing workflow and study its limitations in terms of software availability, dependencies, configuration, portability, and documentation. We rebuild the workflow's software stack to reproduce the published images; we use the software stack for our analysis of discrepancies between original and reproduced results. We document each step in this process, starting from a systematic assessment of the availability of data, software, and documentation. We deliver a collection of fully documented containers for data validation and image reconstruction. Finally, we compile guidelines to increase the reproducibility of computational workflows in scientific projects. Our work enhances the reproducibility and reach of scientific projects like the EHT project, and facilitates the engagement of the overall scientific community, including postdocs and students, regardless of the domain.

\section{M87 Event Horizon Telescope (EHT)}
\label{sec:EHT}

The EHT project uses Very Long Baseline Interferometery (VLBI) to link together eight radio telescopes around the world to study the immediate environment of a black hole with angular resolution comparable to the size of the black hole itself. In April 2019, the EHT Collaboration published measurements of the properties of the central radio source in M87~\cite{collaboration2019first}, including the first direct image of a black hole. The results, that received world wide attention, revealed for the first time a bright ring formed as light bends in the intense gravity around a black hole in the galaxy M87. The black hole is 6.5 billion times larger than the Sun. 

The EHT project provides links to their calibrated data~\cite{eht-dataset} published in CyVerse Data Commons, a publication describing their data processing and calibration~\cite{akiyama2019firstIII}, a link to the software used in their imaging workflow~\cite{eht-code}, and a publication describing the imaging workflow~\cite{akiyama2019firstIV}. The EHT Collaboration released both data products and software, hosting them on third-party repositories. This is a common approach for many NSF-funded projects ranging in size from individual investigators to international collaborations.

\section{Characterization of the EHT Workflow}
\label{sec:Characterization}

\begin{figure*}[!ht]
    \centering
    \includegraphics[width=.8\textwidth]{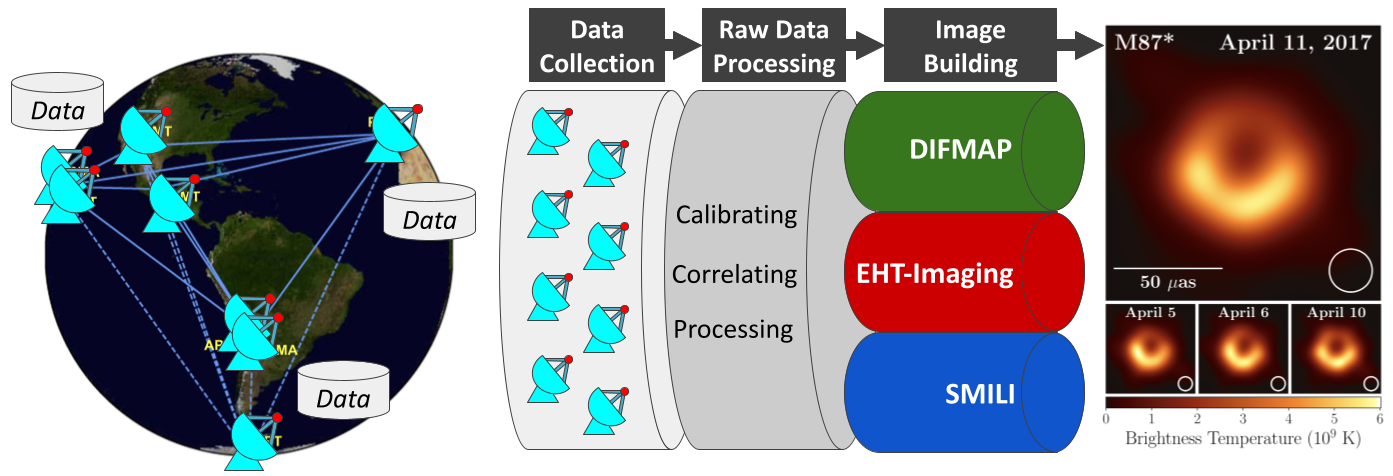}
    \caption{High level overview of the EHT project with its eight telescopes collecting radio interferometry data, its three workflow components including three pipelines for image building, and an image of the M87 black hole extracted from Figure 3 in~\cite{collaboration2019first}.}
    \label{fig:eht-high-level}
\end{figure*}

The EHT workflow comprises three key components: the data collection, the data processing, and image building (see Figure~\ref{fig:eht-high-level}).

{\it Data Collection.} Eight telescopes in the EHT network collect radio interferometry data on certain days at certain times that have permissible weather conditions for all sites, allowing the gathering of data from multiple angles and effectively turning the Earth into one single giant telescope. The EHT data used for the generation of the first M87 black hole images consists of spatiotemporal data of visibility amplitudes collected over five days in 2017 (i.e., April 5, 6, 7, 10, and 11). For each day, collected raw data contains both high and low telescope frequencies.

{\it Raw Data Processing.} Raw data is first pieced together by using the Earth’s geometry and clock/delay model to obtain a common time reference and the pairwise correlation coefficients are computed. 
Then, the data is reduced to a manageable size for use in source imaging and model fitting: data is fringe-fitted, calibrated a priori, and network calibrated. Fringe-fitting is performed using the EHT-HOPS Pipeline for Millimeter VLBI Data Reduction~\cite{blackburn2019eht}. Data undergoes a priori calibration and network calibration in the post-processing stage of the EHT-HOPS pipeline to create \texttt{.uvfits}~\cite{akiyama2019firstIII} files. 
The processed data is stored in the First M87 EHT Results~\cite{eht-dataset} data repository in \texttt{.csv}, \texttt{.txt}, and \texttt{.uvfits} formats and are available to the community. We use this processed data for analysis as  both raw data is not open-access and processing scripts are not open-source at the time of this reproducibility study.

\textit{Image Building.} To reduce biases and increase trust in results, the EHT Collaboration uses three independently-designed pipelines to generate the black hole images. They are: the Difmap M87 Stokes I Imaging Pipeline (DIFMAP)~\cite{difmap_repo}, the EHT-Imaging M87 Stokes I Imaging Pipeline (EHT-Imaging)~\cite{eht-imaging_repo}, and the Sparse Modeling Imaging Library for Interferometry (SMILI)~\cite{smili_repo}. Each pipeline is based on different methods, algorithms, and software libraries but uses the same input data. While the code for each individual pipeline is available as open-source software, the repositories do not contain all of the scripts for image post-processing and generation. Providing documentation for scientific software is challenging and we find that documentation for packaging, installing, and running the pipelines can be incomplete or or is unavailable for certain parts of the analysis.

Table~\ref{tab:availability} lists the available, unavailable, and incomplete data, scripts, code, and documentation used by the EHT workflow and shared with the community before our reproducibility study. To succeed in our effort, we generated and made available the missing components.
\begin{table*}[!ht]
    \centering
    \caption{Availability of data, scripts, code, and documentation before our reproducibility study. Available and incomplete components are linked to the paper presenting them; missing components are marked as unavailable. }
    \label{tab:availability}
    \resizebox{.8\linewidth}{!}{%
    \begin{tabular}{@{}lll@{}l@{\hskip 0.2cm}l@{}}
    \toprule[2pt]
    \multicolumn{5}{@{}l}{\textbf{Data}}                                                                           \\ \midrule[.5pt]
     & Raw data        & \multicolumn{3}{@{}l}{Unavailable} \\     
     & Processed data  & \multicolumn{3}{@{}l}{Available~\cite{eht-dataset}}                                               \\ \midrule[1.5pt]
    \multicolumn{5}{@{}l}{\textbf{Scripts}}                                                           \\ \midrule[.5pt]
      & Raw data processing  & \multicolumn{3}{@{}l}{Unavailable}    \\              
      & Processed data validation & \multicolumn{3}{@{}l}{Unavailable}                                               \\
     & Image post-processing & \multicolumn{3}{@{}l}{Unavailable}                                               \\
     & Figure generation   & \multicolumn{3}{@{}l}{Unavailable}                                               \\ \midrule[1.5pt]
    \multicolumn{2}{@{}l}{\textbf{Pipelines}} & \multicolumn{1}{@{}l}{\textbf{DIFMAP}}     & \textbf{EHT-Imaging} & \textbf{SMILI}      \\ \midrule[.5pt]
     & Code            & \multicolumn{1}{@{}l}{Available~\cite{difmap_repo}}    & Available~\cite{eht-imaging_repo}     & Available~\cite{smili_repo}     \\
     & Packaging       & 
    \multicolumn{1}{@{}l}{Unavailable} & Unavailable & Unavailable  \\
    & Installation manual         & \multicolumn{1}{@{}l}{Incomplete~\cite{difmap_repo}} & Incomplete~\cite{eht-imaging_repo}  & Incomplete~\cite{smili_repo}   \\
    & User manual     & \multicolumn{1}{@{}l}{Incomplete~\cite{difmap_repo}}  & Incomplete~\cite{eht-imaging_repo} & Incomplete~\cite{smili_repo}   \\
     \bottomrule[2pt]
    \end{tabular}%
    }
    \end{table*}

\section{Validating the Data Integrity}\label{sec:DataIntegrity}

A key aspect of any work when reproducing scientific results is the validation of the data integrity: the data used for the generation of the original EHT images should match the data made available to the community. The integrity of data is often considered secondary but can compromise any reproducibilty effort, as it was previously demonstrated by the author in~\cite{brown2021reproducing}. Figure 1 in~\cite{akiyama2019firstIV} characterizes the original data in terms of telescope baselines (i.e., u-v coverages). Scripts to compare the properties of the original data with available data were not available. We generated the missing Python scripts and integrated them into a Jupyter notebook using standard Python modules such as \texttt{matplotlib}, \texttt{pandas}, and \texttt{numpy}.

\begin{figure} [!ht]
\centering
\begin{subfigure}{0.5\textwidth}
  \centering
  \includegraphics[width=\linewidth]{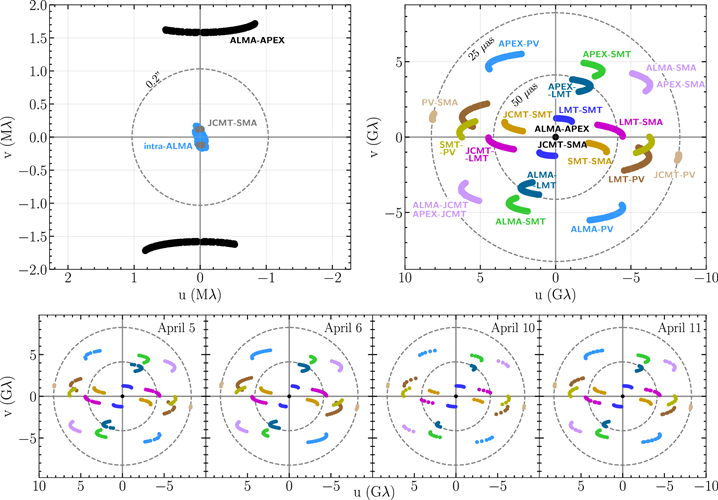}
  \caption{Original}
  \label{fig:sub1}
\end{subfigure}%

\begin{subfigure}{0.5\textwidth}
  \centering
  \includegraphics[width=\linewidth]{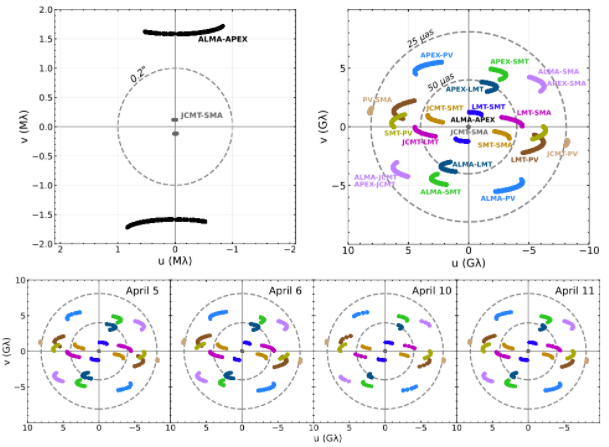}
  \caption{Reproduced}
  \label{fig:sub2}
\end{subfigure}
\caption{Comparison of the properties embedded in the original data used by the the EHT project (from Figure 1 in~\cite{akiyama2019firstIV}) (a) and the available data used in our reproducibility study (b).}
\label{fig:data-validation1}
\end{figure}

Figure~\ref{fig:data-validation1} shows the comparison between the properties of the original data used in Figure 1 in~\cite{akiyama2019firstIV} (set of sub-figures in Fig.~\ref{fig:data-validation1}(a)) and the reproduced properties using the available data (set of sub-figures in Fig.~\ref{fig:data-validation1}(b)). Top left plots in the two sets represent the intra-site EHT interferometer baselines (short baselines). The top right plots represent the aggregate baseline coverage of the EHT array for all four days observed. The bottom plots show the short and long baseline coverage observed by each telescope set at high and low frequencies each day. Qualitatively we can assess the integrity of the data that we input to the three pipelines (i.e., DIFMAP, EHT-imaging, and SMILI). The only difference is the incomplete left plot in Fig.~\ref{fig:data-validation1}(b) due to the fact that the analysis is based on both the available processed EHT data and the unavailable intra-ALMA data from the Atacama Large Millimeter/submillimeter Array (ALMA). This external dataset is not included in the EHT Data Products; based on communications with the EHT Collaboration, the data is not needed for the pipelines to be able to reproduce the black hole images.
\section{Rebuilding the EHT Software Stack}
\label{sec:Rebuilding}

\begin{figure*}[!ht]
\centering
     \scalebox{.32}{\includegraphics{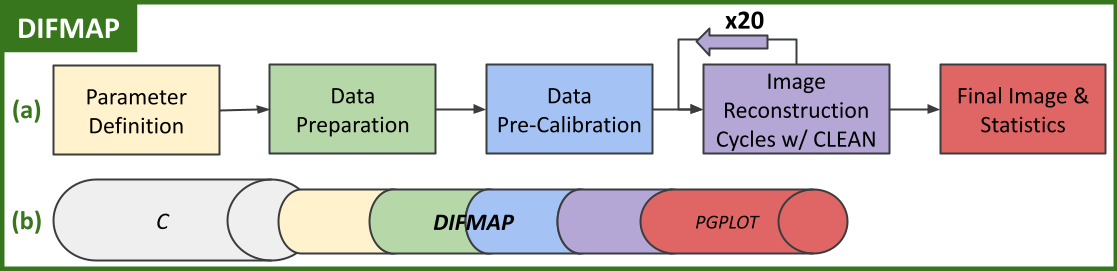}}
     \caption{Detailed view of the DIFMAP pipeline and its implementation.}
   \label{fig:difmap}
\end{figure*}

\begin{figure*}[!ht]
\centering
     \scalebox{.32}{\includegraphics{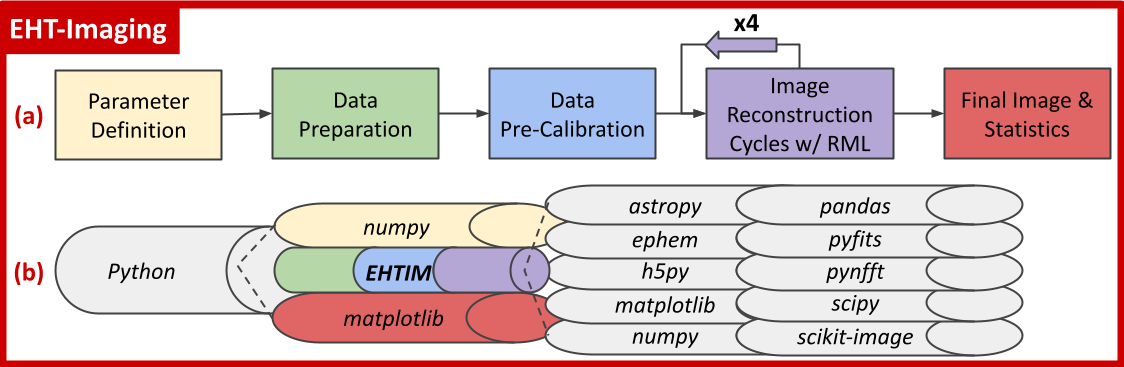}}
     \caption{Detailed view of the EHT-Imaging pipeline and its implementation.}
   \label{fig:eht-imaging}
\end{figure*}

\begin{figure*}[!ht]
\centering
     \scalebox{.32}{\includegraphics{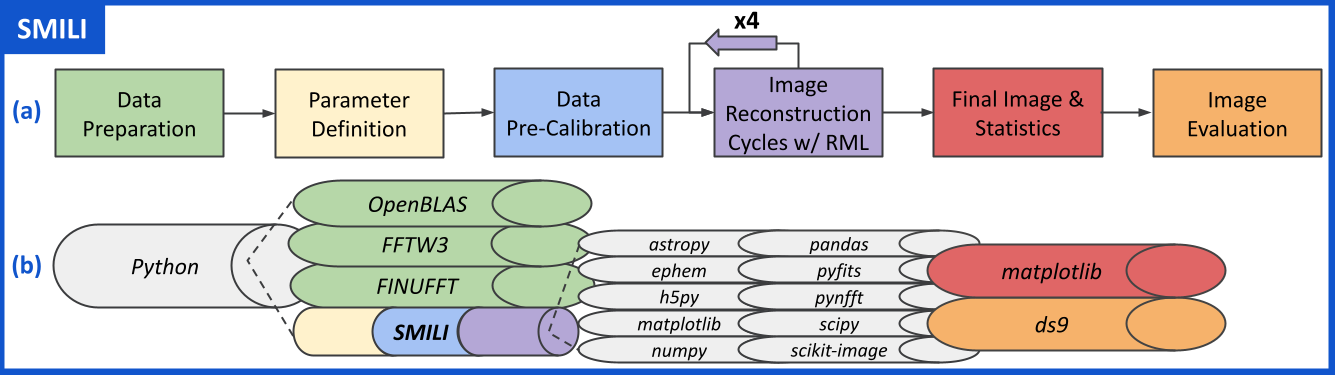}}
     \caption{Detailed view of the SMILI pipeline and its implementation.}
   \label{fig:smili}
\end{figure*}

The three EHT pipelines that are part of image building can be modeled in terms of their functional modules (Figures~\ref{fig:difmap}(a),~\ref{fig:eht-imaging}(a), and~\ref{fig:smili}(a)).
Each pipeline comprises a {\it parameter definition} module for users to establish workflow-specific behaviour as well as {\it data preparation} and \textit{data pre-calibration} modules to pre-process the input files that are fed to the core of each pipeline. A module performing the {\it image reconstruction cycles} runs the image reconstruction algorithm; note how each pipeline uses a different number of cycles. The output of each pipeline includes a {\it final image and statistics} module that is used for qualitative and quantitative analysis of the reconstructed results, respectively. In SMILI, the first two modules are inverted and a \textit{image evaluation} module for data visualization is available at the end of the pipeline. 
 
Although the three pipelines share similar high-level steps, each of them has its own set of auxiliary steps, dependencies, and implementation. Figures~\ref{fig:difmap}(b),~\ref{fig:eht-imaging}(b), and~\ref{fig:smili}(b)) show the dependencies and software components of each pipeline in relation to its functional modules.

{\bf DIFMAP} (Figure~\ref{fig:difmap}) is written in C and uses the CLEAN algorithm for image reconstruction involving iterative deconvolution, paired with a technique called ``difference mapping.'' EHT's DIFMAP script takes a file containing observation data, a mask (set of cleaning windows) file that defines areas of interest for the algorithm to iterate upon, and five command-line arguments, which have been provided in the EHT repository~\cite{eht-code}. After loading this file, the script initializes values, reads the file specifying the mask, and begins the pre-calibration phase, which involves its first cleaning and phase self-calibration. Afterwards, the image undergoes twenty rounds of amplitude self-calibrations and cleanings, and this is when image reconstruction occurs.

{\bf EHT-Imaging} (Figure~\ref{fig:eht-imaging}) uses the Regularized Maximum Likelihood (RML) method of image reconstruction and relies heavily on the \texttt{eht-imaging} Python module (EHTIM) to complete its processes. The EHTIM module defines numerous classes to allow the loading, simulation, and manipulation of VLBI data. By leveraging the classes in this module, the EHT-Imaging workflow loads both the low and high band data files of a single day's observations into a data object and performs various data preparation and pre-calibration steps. The workflow then moves to an imaging cycle with four iterations. Each successive iteration relies directly on the image generated in the previous iteration. After four iterations, the final image is output. The pipeline also allows for optional outputs including the final image and an image summary file containing various imaging parameters and data related to the imaging process.

{\bf SMILI} (Figure~\ref{fig:smili}) is also written in Pyhton and  uses RML like EHT-Imaging. Prior to imaging, SMILI also uses the EHTIM module in order to use data sets pre-calibrated consistently with the other workflows. After the pre-calibration stage, the software generates data tables that are used for the final imaging process. Reconstruction of an image begins with a circular Gaussian with successive iterations relying on the image generated from the previous iteration. There are four stages of iterations with each stage performing three imaging cycles. Once completed, the software outputs the final image and packages the input, pre-calibrated, and self-calibrated data files for traceability. 

Note that each pipeline has its own GitHub repository~\cite{difmap_repo,eht-imaging_repo, smili_repo}. The compilation of each pipeline's original code from the three EHT repositories resulted in several errors. For example, on a Power9 system we missed dependencies and had to remove optimization compilation flags from the installation script to generate the executable code successfully. In general none of the three pipeline codes include a comprehensive list of required software dependencies and libraries used or their versions. We solved dependencies manually by editing problematic scripts; we used Spack, Anaconda, and Pip to install the latest stable version of each necessary library.
Once the compilation was successfully completed, we experienced runtime errors with EHT-Imaging and SMILI that we solved by correcting syntax issues in part of the Python code. We could not find documentation on how to transform the grayscale output of DIFMAP and SMILI into the colored and formatted images from Figure 11 in \cite{akiyama2019firstIV}. We solved this issue by utilizing the EHTIM module for post-processing of grayscale output. 
In the process of rebuilding the EHT software stack, we documented the software packages used, their dependencies, the compilation requirements, and the execution processes for all three pipelines, completing the unavailable or incomplete components in Figure~\ref{tab:availability}.
\section{Packaging and Distribution}
\label{sec:packaging}

To support the portability of the ETH workflow across different platforms, we created a collection of four Docker containers that allows users to reproduce two key results from the EHT project: the characterization of available data (i.e., Figure 1 in~\cite{akiyama2019firstIV}) and the final EHT images of the black hole in Figure 11 in~\cite{akiyama2019firstIV}). 
A first container hosts the entire setting to reproduce the validation of the data integrity; its includes the data tarball from the EHT Data Product page along with our Bash, Python, and Docker scripts. We developed these scripts to automate the installation and configuration of the environment in an easily accessible and portable way. In order for users to be completely satisfied with the validation of the data integrity, we have incorporated a spare tarball within the container for users to perform the \texttt{md5sum} program on it to compare with \texttt{md5sum} of the data from the EHT Data Products page. If both \texttt{md5sum} match, then users knows that the data in the Data Products page has not been modified in any way, and they can move on with the validation by running the Python scripts to reproduce the images of the black hole. The other three containers are used to reproduce the final EHT images of the black hole. Each of the EHT pipelines is packaged into an independent container that automates their installation, dependency setup, environment configuration, and execution. The containers include our own scripts and auxiliary files for conducting the image post-processing steps, which are not available in the original EHT repository. 

All four containers are publicly available in a Docker Hub\footnote{\url{https://hub.docker.com/repository/docker/globalcomputinglab/reproducibility-eht}}. Additional documentation for deploying and using these containers is available in Github\footnote{\url{https://github.com/TauferLab/Reproducibility_EHT}}, along with the scripts to generate the figures reproduced in this paper. These materials augment existing containers in the EHT Docker Hub\footnote{\url{https://hub.docker.com/u/eventhorizontelescope}} and the EHT repositories~\cite{eht-code}.

\section{Reproducing EHT Images}
\label{sec:results}

\begin{figure} [!ht]
\centering
\begin{subfigure}{.5\textwidth}
  \centering
  \includegraphics[width=\linewidth]{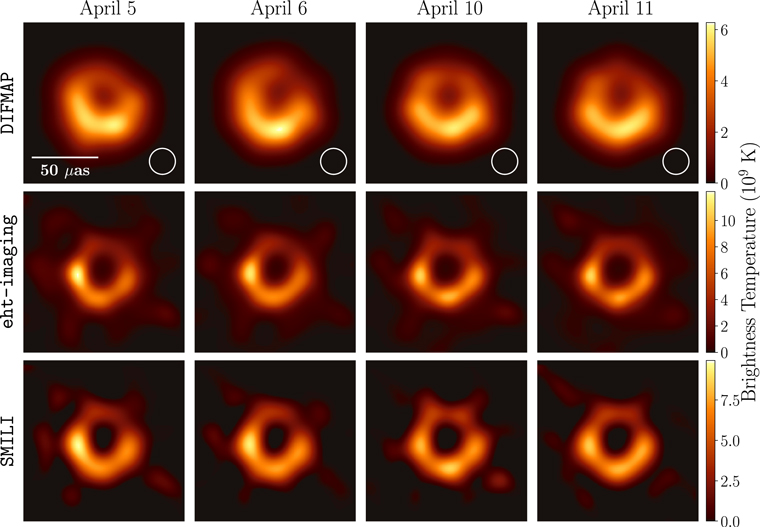}
  \caption{Original}
  \label{fig:original}
\end{subfigure}%

\begin{subfigure}{.5\textwidth}
  \centering
  \includegraphics[width=0.98\linewidth]{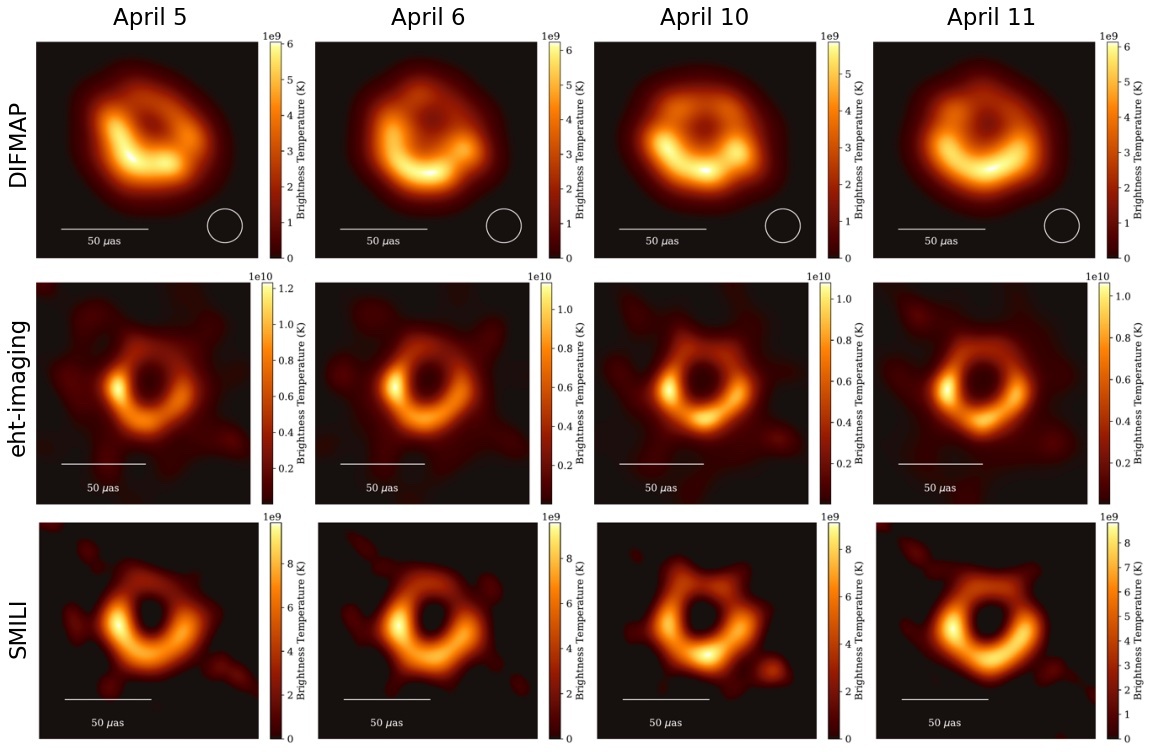}
  \caption{Reproduced}
  \label{final_fig}
\end{subfigure}
\caption{Final images obtained from the original publication from Figure 11 in~\cite{akiyama2019firstIV} (a) and our reproduced results (b).}
\label{fig:images}
\end{figure}

We tested the containerized pipelines both on commodity hardware (a laptop with Inter CPU) and a Power9 cluster at the University of Tennessee, Knoxville. Figure~\ref{fig:images} compares our results: Figure~\ref{fig:original} shows the original images from Figure 11 in~\cite{akiyama2019firstIV} and Figure~\ref{final_fig} shows our reproduced images using the containerized pipelines. The two figures show that we can reproduce the M87 images for all three pipelines.

The images in Figure~\ref{fig:images} provide us with a qualitative comparison. Both sets of images look visually similar in terms of shape and brightness, and the similarity is consistent across pipelines. To perform a quantitative analysis, we compare the ``closure'' quantities reported in Table 5 in~\cite{akiyama2019firstIV} with those reported by our executions of the the three pipelines. For each day and each pipeline, we compare both the $\chi^2_\mathrm{CP}$ and $\chi^2_{\log\mathrm{CA}}$ quantities computed across the top set of parameters~\cite{akiyama2019firstIV}. For brevity, we only report the values with $0\%$ systematic uncertainty. We observe consistency between the two sets of results with no perfect agreement for the EHT-Imaging and SMILI pipelines. We also find a larger difference between the original and reproduced values for the DIFMAP pipeline: this is consistent with the discussion of the different time averaging used in DIFMAP.

\begin{table*}[!ht]
\centering
\begin{tabular}{@{}clcccccc@{}}
\toprule[2pt]
\multicolumn{2}{c}{\multirow{2}{*}{\textbf{}}} & \multicolumn{2}{c}{\textbf{DIFMAP}} & \multicolumn{2}{c}{\textbf{EHT-Imaging}} & \multicolumn{2}{c}{\textbf{SMILI}} \\ \cmidrule(l){3-4} \cmidrule(l){5-6} \cmidrule(l){7-8} 
\multicolumn{1}{l}{}         & \multicolumn{1}{l}{} & Original             & Reproduced       & Original        & Reproduced      & Original        & Reproduced      \\ \midrule[1pt]
\multicolumn{1}{@{}l}{April 5}  &                      & \multicolumn{1}{l}{} &                  &                 &                 &                 &                 \\
Top Set:                     & $\chi_{cp}^{2}$      & $9.40 \pm 3.35$      & $6.81 (\delta=2.59)$  & $1.00 \pm 0.13$ & $1.03 (\delta = 0.03)$ & $1.09 \pm 0.21$ & $1.08 (\delta = 0.01)$ \\
                             & $\chi_{logCA}^{2}$   & $4.99 \pm 1.17$      & $35.12 (\delta = 30.13)$ & $0.97 \pm 0.27$ & $1.38 (\delta = 0.41)$ & $1.11 \pm 0.28$ & $1.29 (\delta = 0.18)$ \\ \midrule
\multicolumn{1}{@{}l}{April 6}  &                      & \multicolumn{1}{l}{} &                  &                 &                 &                 &                 \\
Top Set:                     & $\chi_{cp}^{2}$      & $4.40 \pm 1.45$      & $2.84 (\delta = 1.56)$  & $1.59 \pm 0.16$ & $0.88 (\delta = 0.71)$ & $1.49 \pm 0.23$ & $1.30 (\delta = 0.19)$  \\
                             & $\chi_{logCA}^{2}$   & $3.49 \pm 1.51$      & $24.43 (\delta = 20.94)$ & $1.00 \pm 0.14$ & $1.21 (\delta = 0.21)$ & $1.22 \pm 0.25$ & $1.62 (\delta = 0.40)$ \\ \midrule
\multicolumn{1}{@{}l}{April 10} &                      & \multicolumn{1}{l}{} &                  &                 &                 &                 &                 \\
Top Set:                     & $\chi_{cp}^{2}$      & $3.93 \pm 2.10$      & $2.36 (\delta = 1.57)$  & $0.83 \pm 0.10$ & $0.99 (\delta = 0.16)$ & $0.75 \pm 0.14$ & $1.24 (\delta = 0.49)$ \\
                             & $\chi_{logCA}^{2}$   & $1.57 \pm 0.64$      & $22.74 (\delta = 21.17)$ & $1.30 \pm 0.17$ & $0.82 (\delta = 0.48)$ & $1.11 \pm 0.32$ & $0.86 (\delta = 0.25)$ \\ \midrule
\multicolumn{1}{@{}l}{April 11} &                      & \multicolumn{1}{l}{} &                  &                 &                 &                 &                 \\
Top Set:                     & $\chi_{cp}^{2}$      & $4.01 \pm 1.67$      & $1.91 (\delta = 2.1)$  & $0.97 \pm 0.10$ & $0.88 (\delta = 0.09)$ & $1.23 \pm 0.28$ & $1.00 (\delta = 0.23)$    \\
                             & $\chi_{logCA}^{2}$   & $3.33 \pm 1.01$      & $61.32 (\delta = 57.99)$ & $0.99 \pm 0.13$ & $0.85 (\delta = 0.14)$ & $1.14 \pm 0.13$ & $1.04 (\delta = 0.1)$ \\ \bottomrule[2pt]
\end{tabular}
\caption{Closure quantity $\chi^{2}$ values and statistics for top set images with 0\% systematic uncertainty. We compute the difference $\delta$ between the Top Set values in Table 5 in~\cite{akiyama2019firstIV} and our reproduced values. None of the values agree exactly, but our value is consistent with the spread reported in \cite{akiyama2019firstIV} for the EHT-Imaging and SMILI pipelines.}
\label{tab:closures}
\end{table*}

\section{Lessons Learned and Guidelines}

We compile lessons learned and guidelines to support the reproducibility of scientific projects based on our experience and observations reproducing the M87 black hole images from the EHT project. 

{\bf  Data Availability.} The unavailability of the raw data made the direct validation of the pipeline input data unfeasible. As a proxy for data validation, we reproduced Figure 1 in~\cite{akiyama2019firstIV}, as this figure captures properties of the data telescope frequency and coverage. We are able to reproduce most of these properties except for the intra-site EHT interferometer baselines (short baselines) because the data from the intra-ALMA data) is not available. While this was not the case in this study, any incomplete or missing dataset may result in the users inability to fully verify the data integrity, and can threaten the entire reproducibility process. Data size or ownership constraints can be an obstacle to make raw data available to the public. Under these circumstances, data integrity mechanisms such as hashes ensure the correctness of processed data when releasing the raw data is not feasible. We add the additional service to run an MD5 integrity check for the pipeline input data as part of our EHT container set to facilitate data integrity validation.

{\bf Software Availability.} Several pieces of software were unavailable at different stages of the EHT workflow, and for the three pipelines. The raw data and corresponding software to process the data are not available, neither are the scripts to run the data validation. We developed those scripts for data validation purposes. The code for running the three pipelines is completely available but the image post-processing scripts are not, which forced us to experiment with different settings in order to obtain results comparable to the original for each pipeline. Finally, the plotting libraries used to reproduce the results in Figure 1 from \cite{akiyama2019firstIV} were insufficiently defined. Thus, we manually tuned our plotting scripts to obtain a suitable plotting configuration. The qualitative differences between the original and reproduced images can be the result of our manual tuning, which indicates that just sharing the core for the three pipelines is not sufficient to reproduce the original  images of the black hole in the galaxy M87. To support portability across platforms, we generated four containers that allow users to execute the data integrity validation and original pipeline codes. We also enable execution of the end-to-end workflow by providing all auxiliary materials for the image post-processing, figure generation, and result analysis. 

{\bf Documentation Availability.} In general, there is insufficient documentation on how to package, install, and execute the EHT pipelines, as well as on how to perform both qualitative and quantitative analyses of the results. This hinders the overall reproducibility effort. For instance, documentation is key to reproduce Table 5 in~\cite{akiyama2019firstIV}. Regarding the pipelines, there is insufficient information about software dependencies and versions used as well as file locations and their use. Documenting the whole EHT workflow beyond its image reconstruction components is crucial for the successful reproducibility of the results. Our documentation covering configuration and use of the entire EHT workflow was instrumental for the success of our reproducibility study.

{\bf Software Packaging.} The incomplete documentation resulted in installation, dependency, and portability challenges. We manually edited dependencies to allow installation and compilation, and had to override installation instructions that were resulting in unstable environments. We found that by containerizing the workflow we can hide these challenges from the end user, simplifying the installation and deployment of the EHT workflow.

{\bf Methods Description.} Incomplete descriptions of the results analysis process (e.g., the data averaging time or the additional systematic error budget added to the uncertainties) complicates the reproducibility of the $\chi^{2}$ statistics in Table 5 in~\cite{akiyama2019firstIV}, as errors add up quickly. Conducting an adequate quantitative assessment of the final results becomes very challenging under these circumstances. Members of the EHT Collaboration highlighted in conversations with the authors of this paper how a qualitative comparison of the images is more interpretable.

{\bf Access to Final Results.} The authors of~\cite{akiyama2019firstIV} did not release the output fiducial images, and therefore we did not have a fixed reference to use for direct comparison of our reproduced images and the original ones. This, in addition to partial access to the data and incomplete description of the methods used, prevents us from conducting a complete validation of the reproduced images.

{\bf Access to Distributed Knowledge.} The EHT Collaboration made substantial investments to allow independent users to qualitative and quantitatively reproduce their results and ensure the robustness of the EHT project. Nonetheless, we found challenging to reproduce the original results without direct knowledge of the methods and analyses, or the direct collaboration with the authors of the studies. Our experience illustrates the general challenges that users external to a project face when gathering knowledge on data, code, and documentation that was originally generated from multiple teams in a distributed fashion. The effort of the EHT Collaboration to remove biases by designing and deploying three completely separate pipelines, while instrumental for the trustworthiness of the project results, is also an obstacle to the project reproducibilty. 

\section{Conclusions}

In this paper we deliver our experience reproducing the black hole images from the EHT project, and report new guidance and practices for building reproducible scientific research. Our work complements the work of the EHT Collaboration with supplemental data, scripts, documentation, and a set of containers. Postdocs, graduate and undergraduate students, and even high school students can benefit from accessing our data and code, and using our documentation to reproduce findings from the EHT project, learn about the EHT funding, and ultimately get involved in STEM research. Our guidance and practices can be incorporated more broadly by other scientific workflows. The EHT project continues to be a leader in reproducibility efforts and have provided comprehensive data products for their recent observations of Sag A$^\ast$.

Assessing the level of detail required to cover the vast knowledge developed in a project the size of EHT is a complex task. Finding the balance between the effort from original research teams to enable reproducibility, and users attempting to reproduce the results is still an open question. Our experience with the EHT and LIGO projects reveals an important and recurring issue in reproducibility: challenges remain in disseminating findings in a way that allows reproducibility of results without direct interaction with the original team that produced them.

\section{ACKNOWLEDGMENT}
This work was supported in part by NSF grants \#2041977, \#1941443, \#2041901, \#1664162 and \#2041878. We thank Kazunori Akiyama, Lindy Blackburn, Chi-kwan Chan, and Rebecca White for helpful discussions.

\begin{IEEEbiography}{Ria Patel}{\,}is a fourth-year undergraduate student majoring in Computer Science and minoring in Mathematics and Physics at the University of Tennessee, Knoxville. She is an Undergraduate Research Assistant in the Global Computing Lab under Dr. Taufer. 
\end{IEEEbiography}

\begin{IEEEbiography}{Brandan Roachell}{\,}is an undergraduate student majoring in Computer Science and minoring in Mathematics at the University of Tennessee, Knoxville. He is an Undergraduate Research Assistant in the Global Computing Lab under Dr. Taufer. 
\end{IEEEbiography}

\begin{IEEEbiography}
{Silvina Caíno-Lores}
is a Research Assistant Professor at the University of Tennessee, Knoxville. She obtained her Ph.D. in Computer Science and Technology from Carlos III University of Madrid (Spain) in 2019. Her research interests include cloud computing, in-memory computing and storage, HPC scientific simulations, and data-centric paradigms.

\end{IEEEbiography}

\begin{IEEEbiography}{Ross Ketron}{\,}is an undergraduate student  in Computer Science at the University of Tennessee, Knoxville and an Undergraduate Research Assistant in the Global Computing Lab under Dr. Taufer. 
\end{IEEEbiography}

\begin{IEEEbiography}{Jacob Leonard}{\,}is an undergraduate student  in Computer Science at the University of Tennessee, Knoxville and an Undergraduate Research Assistant in the Global Computing Lab under Dr. Taufer. 
\end{IEEEbiography}

\begin{IEEEbiography}{Nigel Tan}{\,}is an graduate student  in Computer Science at the University of Tennessee, Knoxville. His research interests are in software portability across heterogeneous platforms.
\end{IEEEbiography}

\begin{IEEEbiography}{Karan Vahi}{\,}is a Senior Computer Scientist at USC Information Sciences Institute. Vahi received a M.S in Computer Science in 2003 from University of Southern California. 
His research interests include scientific workflows and distributed computing systems. Contact him at vahi@isi.edu.
\end{IEEEbiography}

\begin{IEEEbiography}{Duncan A. Brown}{\,} is the Charles Brightman Professor of Physics at Syracuse University. Dr.~Brown received a Ph.D. degree in physics from the University of Wisconsin-Milwaukee in 2004. He was a member of the LIGO Scientific Collaboration from 1999 to 2018 and is a fellow of the American Physical Society. 
His research is in gravitational-wave astronomy and astrophysics, and the use of large-scale scientific workflows. Contact him at dabrown@syr.edu.
\end{IEEEbiography}

\begin{IEEEbiography}{Ewa Deelman}{\,} received her PhD in Computer Science from the Rensselaer Polytechnic Institute. She is a Research Director at USC/ISI and a Research Professor at the USC Computer Science Department and the lead the design and development of the Pegasus Workflow Management software.  Her research explores the interplay between automation and the management of scientific workflows that include resource provisioning and data management. 
\end{IEEEbiography}

\begin{IEEEbiography}{Michela Taufer}{\,} holds the Jack Dongarra Professorship in High Performance Computing within the Department of Electrical Engineering and Computer Science at the University of Tennessee, Knoxville. Dr. Taufer received her Ph.D. in computer science from the Swiss Federal Institute of Technology (ETH) in 2002. 
Her interdisciplinary research is at the intersection of computational sciences, high permanence computing and data analytics. 
\end{IEEEbiography}

\bibliographystyle{IEEEtran}
\bibliography{references}
\end{document}